\documentclass[preprint]{aastex}
\begin{document}
\title{Lyman-$\alpha$ Observations of Astrospheres}
\author{Jeffrey L. Linsky}
\affil{JILA, University of Colorado and NIST, Boulder CO, USA}
\email{jlinsky@jila.colorado.edu}
\author{Brian E. Wood}
\affil{US Naval Research Laboratory, Washington DC, USA}
\email{brian.wood@nrl.navy.mil}

\begin{abstract}

Charge-exchange reactions between outflowing stellar wind protons and 
interstellar neutral
hydrogen atoms entering a stellar astrosphere produce a region of
piled-up-decelerated neutral hydrogen called the hydrogen wall. Absorption by
this gas, which is observed in stellar Lyman-$\alpha$ emission lines, 
provides the only viable technique at this time 
for measuring the mass-loss rates of 
F--M dwarf stars. We describe this technique, present an alternative way 
for understanding the relation of 
mass-loss rate with X-ray emission, and identify several critical issues.

\end{abstract}

\section{Introduction}

The measurement of mass-loss rates for F--M dwarf stars and 
the development of theoretical models
to predict these rates have challenged observers and theoreticians for many 
years. While {\em in situ} measurements of the solar-wind mass flux 
have been available since the early space age, such measurements are not 
feasible for other stars. Instead, 
\citet{Wood2005a,Wood2005b} showed that astrospheric absorption in the 
stellar Lyman-$\alpha$ emission line provides a sensitive tool for 
measuring stellar mass-loss rates of late-type dwarf stars. Although other 
techniques have been proposed, including 
the search for charge-exchange-induced X-ray emission \citep{Wargelin2001} 
and free-free emission at radio wavelengths 
\citep{Gaidos2000,Fichtinger2013}, the 
Lyman-$\alpha$ astrospheres technique is presently 
the only technique that provides credible results. 
We describe this technique in Section~2, present an
alternative way for explaining the empirical relation of mass loss to 
X-ray flux in Section~3, and identify some critical issues in 
Section~4. 

\section{Lyman-$\alpha$ Observations of Astrospheres}

The interaction of ionized stellar-wind plasma with inflowing, mostly neutral, 
interstellar gas leads to a rich phenomenology. Figure 1 shows an example of 
a hydrodynamic model of the heliosphere computed by \citet{Muller2004}. 
The salient 
features of this model and other models of the heliosphere and astrospheres 
is the presence of a termination shock where the solar wind becomes subsonic, 
a heliopause/astropause that separates ionized solar/stellar wind plasma from 
charge-exchanged local interstellar medium (LISM) ions,
and the so-called hydrogen wall where inflowing neutral hydrogen gas
piles up, is decelerated, and is heated by the charge-exchange reactions. IBEX 
results \citep{McComas2012} indicate that the inflow velocity of the LISM 
is likely too slow to produce a bow shock, however \citet{Scherer2014}
argue that inclusion of inflowing He$^+$ ions yields Alfv\'en and fast
magnetosonic speeds slower than the inflow speed and thus a bow shock.
As first predicted by \citet{Baranov1993}, hydrogen walls provide the 
observational basis for measuring stellar mass-loss rates. However, the 
small hydrogen column densities in these astrosphere regions, 
$\log N(H I)=$14.0--15.0, means that optical depths in neutral metal 
lines also formed in these regions will be too small to detect, 
leaving Lyman-$\alpha$ as the only 
available spectral diagnostic for measuring stellar mass-loss rates.

Figure 2 shows the effect of neutral hydrogen absorption from 
the LISM and in the hydrogen walls of the star and 
the Sun on the Lyman-$\alpha$ emission line emitted by a star. 
Since the hydrogen wall gas is decelerated relative to the LISM gas,
the solar hydrogen wall appears as extra absorption on the red side of the
LISM absorption, and the stellar hydrogen wall viewed from outside of the star
appears as extra absorption on the blue side of the LISM absorption. Figure~3 
shows the effects of increasing stellar mass-loss rate on
the stellar hydrogen-wall absorption computed with hydrodynamic 
astrosphere models \citep{Wood2005a}. 
This technique has been successfully applied to 15 stars
or binary systems that are relatively close ($d<32$~pc), with relatively small
LISM neutral hydrogen column densities [$\log N(H I)<18.5$]. Clearly,
the LISM gas must contain a significant amount of neutral hydrogen 
for the development of a hydrogen wall and for this technique to work. 
Lyman-$\alpha$ observations of more stars are needed, but 
observational limitations will restrict the number of stellar winds that could 
be measured by this technique.

\section{An Alternative Way for Explaining the Relationship of Mass loss and 
X-ray Flux}

\citet{Wood2005a} proposed that the mass-loss rate per unit stellar
surface area ($\dot{M}$) measurements for G and K dwarf 
stars shown in Fig.~4 can be fit by a power law, 
$\dot{M} \propto F_X^{1.34\pm0.18}$, for X-ray surface fluxes
$F_X < 10^6$ erg~cm$^{-2}$~s$^{-1}$. While the fit looks good considering
the roughly factor of two uncertainties in the mass-loss rate
measurements and the varying 
X-ray fluxs that were measured at different times, there
is no accepted explanation for this power-law relation. We therefore 
propose an alternative way of characterizing these data that suggests
a physical scenario that could explain these results. We start by 
separating the data into three groups rather than by assuming  
a power-law relation.

Group 1 includes the Sun and four other stars (61 Vir, the unresolved
binary $\alpha$~Cen~AB, $\epsilon$~Ind, and 61~Cyg~A). 
All of these stars are slow rotators (22--42
day periods), most have solar-like mass-loss rates (per unit surface 
area), $\dot{M}/\dot{M_{\odot}} \approx 1.0$, 
and $F_X$ in the range $10^4$--$10^5$ ergs~cm$^{-2}$~s$^{-1}$. 
\citet{Cohen2011}, \citet{Wang2010}, and others have shown that the solar 
mass-loss rate is uncorrelated
with $F_X$, although both $\dot{M}$ and $F_X$ vary
by over a factor of 10. The average solar mass loss rate
($\dot{M_{\odot}}\approx 2\times 10^{-14} M_{\odot}$ yr$^{-1}$),
however, is independent of $F_X$  
despite the very different number of active regions on the solar disk 
at times of solar activity minima and maxima. 
 
Group 2 consists of two stars and two unresolved binary systems 
with $F_X$ in the narrow range 3--5$\times 10^5$ ergs~cm$^{-2}$~s$^{-1}$. 
The two stars ($\epsilon$~Eri and $\xi$~Boo~B) are moderate speed 
rotators ($P_{\rm rot}$=11.68 and 11.5 days, respectively). 
The other two targets are unresolved binaries (70~Oph AB and
36~Oph AB) with rotational periods in the range 19.7--22.9 days and
measured mass-loss rates only for the binary systems rather than for 
the individual stars. 

The third group of G--K dwarf stars consists of $\pi^1$~UMa \citep{Wood2014}
and $\xi$~Boo~A \citep{Wood2010}. These stars are rapid rotators 
($P_{\rm rot}$ = 4.69 and 6.2 days, respectively), 
have large $F_X \approx 2\times 10^6$ 
ergs~cm$^{-2}$~s$^{-1}$, but small mass-loss rates similar to the Sun.
We note that these mass-loss rates are a factor of 100 times smaller than
theoretical estimates for $\pi^1$~UMa \citep{Cranmer2011,Drake2013} and for 
$\tau$~Boo, an F7~V star with $P_{\rm rot}=3.0$ days \citep{Vidotto2012}.

The very different behavior between mass loss, which emerges only along open
magnetic-field lines, and X-ray flux, which is emitted mostly in closed
magnetic-field regions with high electron densities, likely depends on the 
energy available for driving mass loss at the coronal base of open field 
lines, the number of open field lines, and the divergence of the open field 
lines with height in the corona. While detailed theoretical studies of wind 
acceleration mechanisms \citep[e.g.,][]{Holzwarth2007,Cranmer2011} and coronal 
magnetic-field morphologies \citep[e.g.,][]{Jardine2013} are needed,
we outline here a simple physical model that may explain the complex 
relationship between $\dot{M}$ and $F_X$ shown in Fig.~4.

\citet{Wang2010} showed that for the 
Sun, the mass flux along an open magnetic-field line is proportional to 
the magnetic-field strength at its coronal base footpoint.
Since the coronal base magnetic-field strength near active 
regions is much larger than for high-latitude coronal holes, 
the mass flux in field lines located near active regions is much larger than
for open field regions away from active regions (e.g., polar holes). 
However, the divergence of field lines emerging 
from near active regions is much larger than for open field 
regions away from active regions. As a result, the greater divergence of 
field lines from active regions compensates for the larger mass flux 
in these field lines leading to the same average solar 
mass-loss rate even when the 
surface coverage of active regions changes as indicated by a
factor of 10 range in $F_X$.
This explanation for the nearly constant value of $\dot{M}$ over at least 
a factor of 10 range in $F_X$ seen in the solar data likely also explains the
results for the other Group 1 stars.

For the faster rotating stars in Group 2, especially $\epsilon$~Eri and 
$\xi$~Boo~B, the much stronger magnetic dynamo produces much larger 
magnetic fluxes \citep{Reiners2012}. This likely means that more active 
regions are distributed across their stellar surfaces
and stronger magnetic-field strengths outside of these active regions
are present in the Group 2 stars compared with the Group 1 stars. 
As a result, the coronal surface area not covered by active regions is 
decreased substantially such that there is much less available volume for 
open field lines to expand. The smaller divergence of field lines from 
active regions will not be able to compensate for the enhanced mass flux along 
these stronger magnetic-field lines. \citet{Cohen2011} showed that in the 
absence of significant magnetic-field divergence, stellar
mass-loss rates will increase with the magnetic field strength and the 
available energy at the base of the magnetic field lines.

A viable explanation for the solar-like mass-loss rates for the 
rapidly rotating stars in Group 3 is a matter of speculation at this time. 
We suggest that the very strong dynamos in these stars produce complex 
magnetic-field morphologies including a strong toroidal component seen
in rapidly rotating stars like $\xi$~Boo~A \citep{Petit2005}. 
These very strong magnetic 
fields are likely closed in very active stellar coronae leaving only a few   
open-field lines from which the stellar wind can emerge. The fewer 
open-field lines could more than 
compensate for the stronger mass-flux rates in the remaining open-field lines.
 
Simulations by \citet{Suzuki2013} of MHD wave propagation upwards
along open magnetic field lines from the photospheres of solar-like stars
may explain the empirical correlation of $\dot{M}$ with $F_X$
shown in Figure~4. They found that with increasing magnetic field
strength and turbulence in the photosphere (the origin of stellar activity), 
the reflection of upwardly propagating Alfv\'en waves is less
efficient, leading to an increase in the energy available in the
corona for mass loss and X-ray emission. However, radiative losses in
the chromosphere increase even faster than the mass-loss flux, 
leading first to $\dot{M}$ saturation at a high level 
and then to decreasing $\dot{M}$ with
increasing $F_X$. Qualitatively, this may explain the increase
and then decrease of $\dot{M}$ with increasing $F_X$, but the
peak $\dot{M}$ corresponding to saturation in these 
calculations is much larger than is observed and the decrease in 
$\dot{M}$ at very large $F_X$ levels was not modeled in detail.

\section{Critical Issues}

The study of cool dwarf star astrospheres has made important progress 
from the early theoretical models of the heliosphere that first showed the
presence of a hydrogen wall \citep{Baranov1993}, through the development of
sophisticated-hydrodynamic models of stellar atmospheres by M\"uller, 
Izmodenov, Zank, and others, to the testing and refinement of these models
by comparison with stellar Lyman-$\alpha$ observations. Despite this
important progress, there remain a number of critical problems requiring 
future theoretical and observational studies:

\begin{description}
\item[{\bf Charge-exchange rates}] The existence and physical properties of  
solar/stellar hydrogen walls depend critically on charge-exchange 
reaction rates between outflowing solar/stellar wind
ions (primarily protons) and the inflowing LISM neutral hydrogen atoms. If the 
presently assumed charge-exchange reactions rates are in error, then
the inferred values for N(H I), heating, and deceleration of neutral hydrogen 
atoms in the hydrogen wall will also be in error. It is important 
to compute charge-exchange reaction rates for the low densities and 
likely non-Maxwellian velocity distributions of the interacting species.

\item[{\bf Proper inclusion of neutrals}] Neutral hydrogen and helium atoms
inflowing from the LISM have long mean-free paths that are not easily 
included in hydrodynamic 
model calculations. The neutrals must be treated with kinetic equations or
Monte Carlo simulations that must be coupled somehow to the hydrodynamic
calculations for the plasma.

\item[{\bf Theoretical models of stellar astrospheres may not include 
all physical processes}] For example, most models and 
simulations assume that
stellar mass loss is constant or quasi-steady state and,
therefore, do not include
transient events like coronal mass ejections (CMEs) that are
seen on the Sun but not yet detected on stars. \citet{Drake2013} argued
that CMEs in active stars could dominate the mass-loss rate based on an
extrapolation of solar data, but this process must be tested in
realistic models and against stellar CME measurements when they become
available. The magnetic field morphology, in particular the height at
which closed field lines open, plays an important role in mass
loss. Realistic models should compute the magnetic field structure
self-consistently with the mass outflow. 

\item[{\bf Accurate boundary conditions}] Whether the LISM gas flowing 
into the heliosphere forms a bow shock or a bow wave requires accurate 
values for the interstellar gas flow speed and magnetic field
strength. Even for the Sun, for which we have the best data, these
quantities are not yet known accurately. 
In particular, the magnetic field strength
outside of the heliopause has not yet been measured by the Voyager
spacecraft or reliably
estimated. For nearby stars, the flow speed of the star with respect
to the interstellar medium gas is probably uncertain by less than 3
km~s$^{-1}$, but the interstellar magnetic field is not well known. Thus
the presence or absence of a bow shock is not known in many cases. A
critical question is whether the interstellar gas flowing into
an astrosphere is partially ionized like the LIC or fully ionized. In
the latter case, there will be an astrosphere but no hydrogen wall and
thus no possibility for measuring the mass-loss rate. The presence of
Lyman-$\alpha$ absorption on the short wavelength side of the interstellar
absorption indicates the presence of a hydrogen wall and that the star
is surrounded by partially ionized interstellar gas, but the inferred value
of $\dot{M}$ depends on the neutral hydrogen number density of the
inflowing gas which has been assumed to be the same as in the LIC given the
absence of other measurements. The nondetection of astrospheric Lyman-$\alpha$
absorption can be explained either by assuming that the star is
embedded in fully-ionized interstellar gas or that the star has an
extremely low mass-loss rate. The latter is unlikely but cannot be
ruled out at this time. 
 
\item[{\bf Active stars appear to have low mass-loss rates}] 
We have suggested that the low mass-loss rates for the two
rapidly rotating stars with $F_X>10^6$ ergs~cm$^{-2}$~s$^{-1}$ result from 
very strong magnetic fields with complex geometries that severely restrict 
the number of open field lines in the corona. A new generation of theoretical
models is needed to test and develop this scenario. There are models 
for very rapidly rotating stars in which magnetocentrifugal acceleration
is important \citep{Vidotto2011} and for T Tauri stars for which accretion
and disk winds are important \citep[e.g.,][]{Vidotto2010}. What is lacking
are models for stars like $\pi^1$ Ori and $\xi$~Boo A, which are active stars
without extreme rotation or T Tauri-like phenomena. Numerical
simulations of MHD wave propagation in open flux tubes embedded in
realistic atmospheres such as those of \citet{Suzuki2013}, 
\citet{Cranmer2011}, and \citet{Holzwarth2007} provide
insight into the essential physics and need to be developed further.
Observations of more stars, especially those with $F_X$ near and above 
$10^6$ erg~cm$^{-2}$~s$^{-1}$, are also 
needed to determine whether the decrease in $\dot{M}$ for 
$F_X \geq10^6$ erg~cm$^{-2}$~s$^{-1}$ is sharp or gradual, or indeed typical. 
Also, reobservations of stellar Lyman-$\alpha$ lines 5--10 
years after the original observations would test whether astrospheres 
and mass-loss rates change during the time scale for stellar winds to 
propagate well out into their astrospheres. 

\end{description}

\section{Conclusions}

The study of astrospheres is intimately connected with the measurements of
stellar-wind mass fluxes for late-type stars. At present the Lyman-$\alpha$
astrospheres technique is the only successful technique for infering 
these mass-loss rates. New theoretical models for stellar astrospheres and 
stellar mass loss are needed to explain in physical terms the results 
provided by this technique. 

\begin{acknowledgements}

JL acknowledges support by the Space Telescope Science Institute
(STScI) to the University of Colorado for HST observing program
GO-11687. BW acknowledges support by STScI for observing program GO-12596.
STScI is operated by the Association of Universities for Research in
Astronomy, Inc. under NASA contract NAS~5-26555.
We thank the referees for their very useful comments.

\end{acknowledgements}

\begin{figure}
\vspace*{-30mm}
\begin{center}
\includegraphics[width=15.0cm]{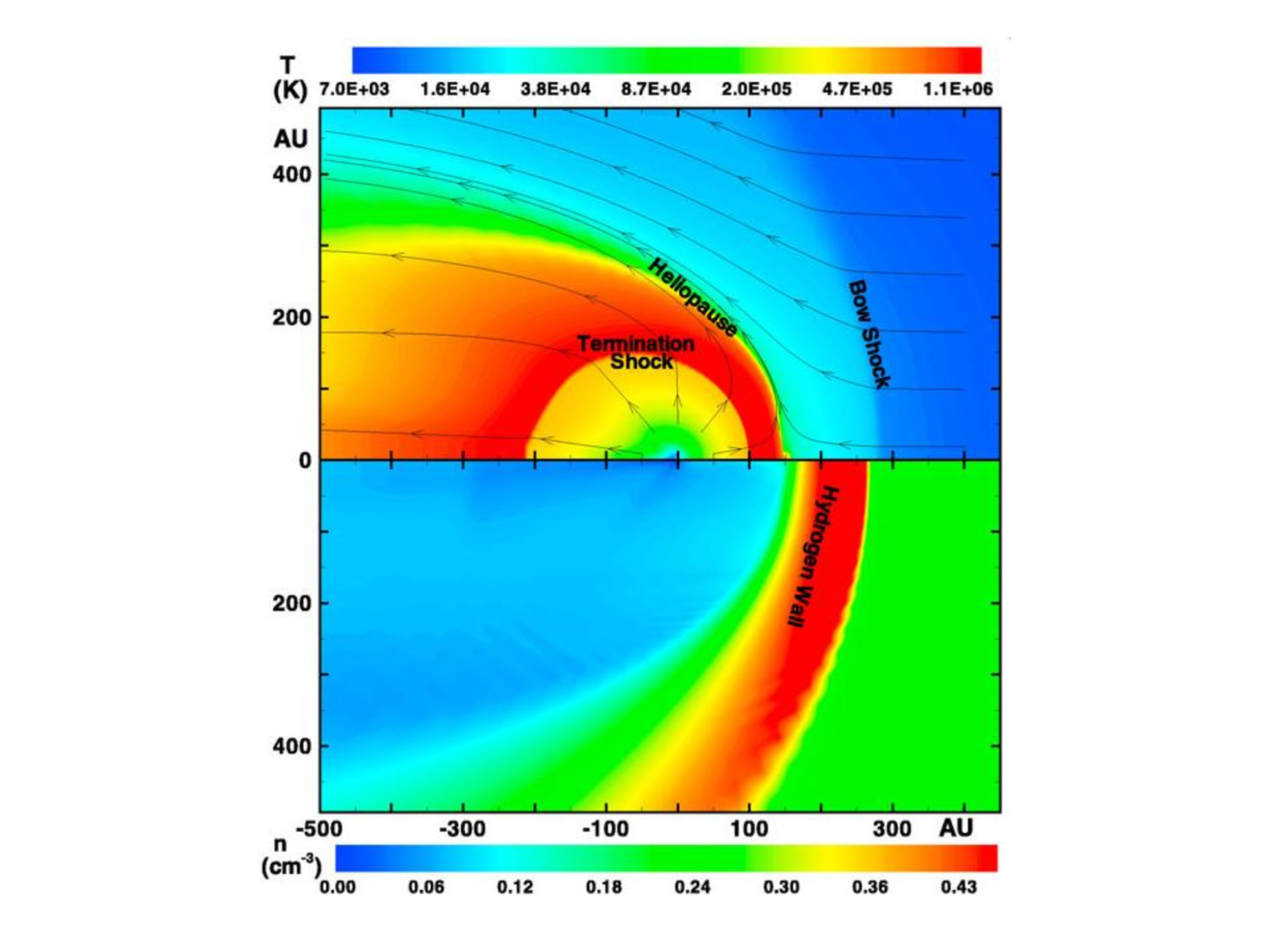}
\end{center}
\caption{This figure shows the different components of a 
heliosphere/astrosphere as seen 
in temperature (top) and neutral hydrogen density (bottom). At the termination 
shock, the supersonic solar/stellar wind becomes subsonic and heats 
the postshock gas. Further out from the Sun/star, the heliopause/astropause 
separates the outflowing-ionized solar/stellar wind gas from the 
interstellar gas that has 
been ionized by charge exchange reactions. 
The pile-up of heated and decelerated neutral hydrogen gas at larger 
distances also results from charge exchange reactions with the outflowing 
solar/stellar 
wind ions. Beyond the hydrogen wall, there may be a bow shock
depending on the inflow speed of interstellar gas and its magnetic 
field strength. Figure from \citet{Muller2004}.}
\end{figure}

\begin{figure}
\vspace*{2mm}
\begin{center}
\includegraphics[width=15.0cm]{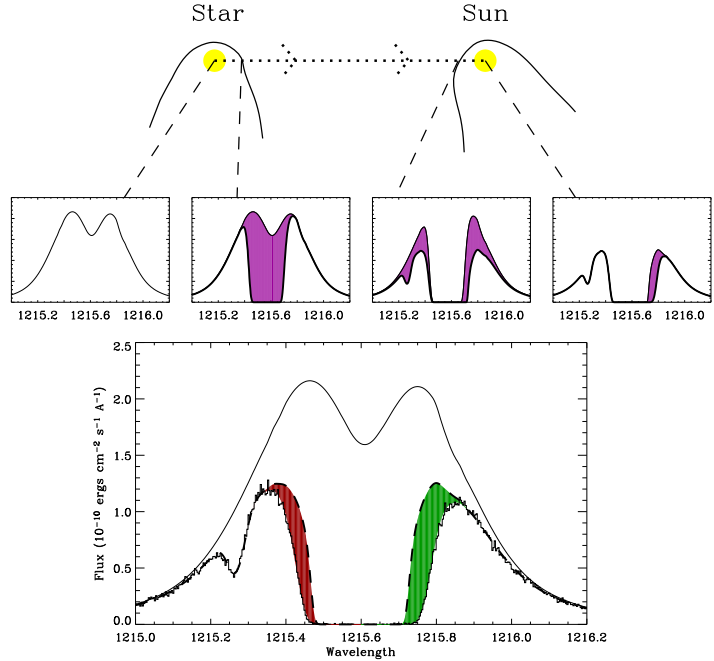}
\end{center}
\caption{The presence of absorption components in the observed 
stellar Lyman-$\alpha$ emission line (bottom) is illustrated in this 
schematic sketch of the journey of Lyman-$\alpha$ photons
from the star to an observer inside the heliosphere \citep{Wood2004}.
Astrospheric absorption is blue-shifted relative to the line of sight 
component of the interstellar flow velocity because the stellar 
hydrogen-wall gas is decelerated (in the stellar reference frame) 
relative to the interstellar medium flow. Lyman-$\alpha$
photons then traverse the LISM and are absorbed by interstellar 
neutral hydrogen at the
projected flow velocity of the LISM. The hydrogen wall gas in the heliosphere
produces redshifted absorption as a result of its deceleration relative to the
interstellar gas. The effect of these three absorption components 
(illustrated in the middle plots) is 
to produce extra absorption
on the blue side of the LISM absorption indicating the presence of a 
hydrogen wall surrounding the star, and absorption on the red side of the LISM
feature produced by the solar hydrogen wall.} 

\end{figure}
\begin{figure}
\vspace*{2mm}
\begin{center}
\includegraphics[width=15.0cm]{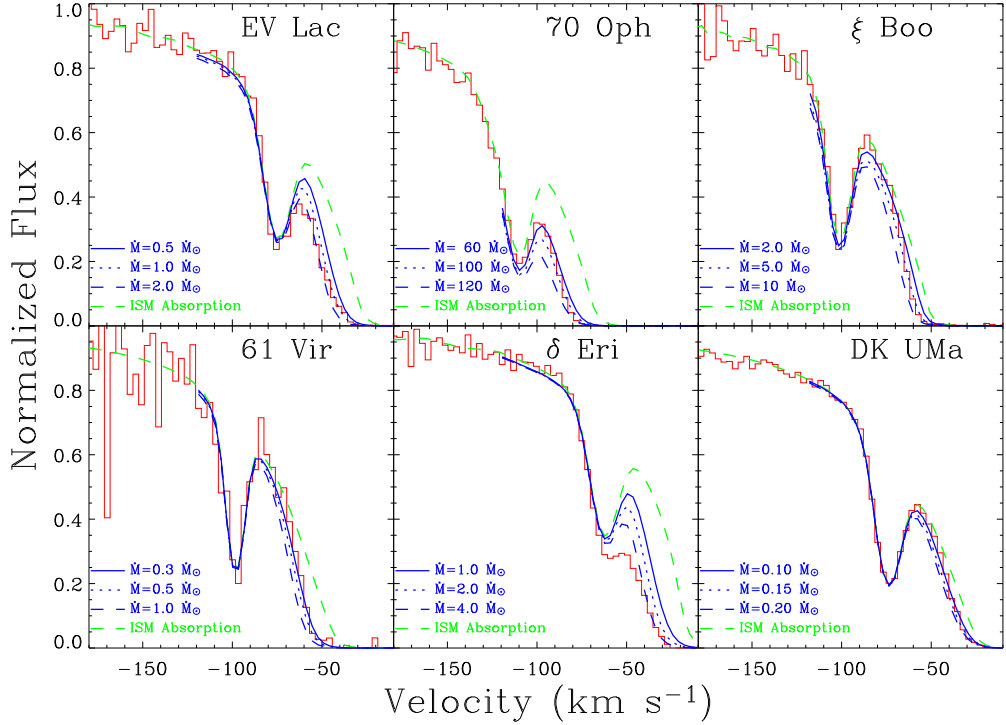}
\end{center}
\caption{Comparisons of the observed absorption on the short wavelength side 
of the Lyman-$\alpha$ lines of six stars with 
hydrodynamic models assuming wind speeds of 400 km~s$^{-1}$ and parameters 
of the LISM \citep{Wood2005a}. The effect of increasing stellar-wind mass 
flux is to increase the amount of blue-shifted absorption by the stellar
hydrogen wall (in units of the solar mass loss rate $\dot{M}=2\times 10^{-14}$ 
solar mass per year). The absorption line centered near -80 km~s$^{-1}$ is
deuterium Lyman-$\alpha$ absorption in the LISM.}
\end{figure}

\begin{figure}
\vspace*{2mm}
\begin{center}
\includegraphics[width=12.0cm]{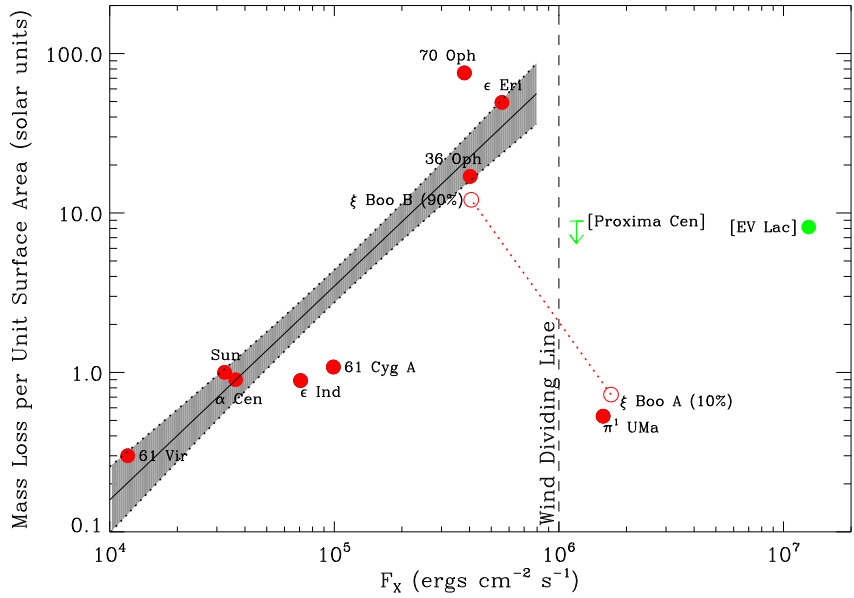}
\end{center}
\caption{A plot of mass loss rate with respect to x-ray flux, both quantities 
expressed per unit stellar 
surface area. For stars with x-ray surface fluxes less than 
$10^6$ erg~cm$^{-2}$~s$^{-1}$, there is a power law relation comparing 
mass flux rate to x-ray flux. This simple relation does not hold for stars 
with larger $F_X$ including the young star $\pi^1$~UMa,
$\xi$~Boo~A, and the M dwarf flare stars Proxima Centauri and EV Lac 
\citep{Wood2014}.} 
\end{figure}

\end{document}